# THE SARPTICAL DATASET FOR JOINT ANALYSIS OF SAR AND OPTICAL IMAGE IN DENSE URBAN AREA

*Yuanyuan Wang*[(1)], *Xiao Xiang Zhu*[(1,2)]

(1) Signal Processing in Earth Observation, Technical University of Munich
(2) Remote Sensing Technology Institute, German Aerospace Center

**ABSTRACT**

The joint interpretation of very high resolution SAR and optical images in dense urban area are not trivial due to the distinct imaging geometry of the two types of images. Especially, the inevitable layover caused by the side-looking SAR imaging geometry renders this task even more challenging. Only until recently, the "SARptical" framework [1], [2] proposed a promising solution to tackle this. SARptical can trace individual SAR scatterers in corresponding high-resolution optical images, via rigorous 3-D reconstruction and matching. This paper introduces the *SARptical dataset*, which is a dataset of over 10,000 pairs of corresponding SAR, and optical image patches extracted from TerraSAR-X high-resolution spotlight images and aerial UltraCAM optical images. This dataset opens new opportunities of multisensory data analysis. One can analyze the geometry, material, and other properties of the imaged object in both SAR and optical image domain. More advanced applications such as SAR and optical image matching via deep learning [3] is now also possible.

## 1. INTRODUCTION

With the growing attention on very high resolution SAR data in urban areas, the fusion of optical and SAR images in dense urban area has become an emerging and timely topic, because the complementary of these two data types can lead us to unprecedented insights and findings, such as the unique scattering mechanisms of different urban infrastructures. Lying at the basis of such fusion task is the challenging co-registration of SAR and optical images. These two types of images are acquired with intrinsically different imaging geometries, and thus are nearly impossible to be co-registered without a precise 3-D model of the imaged scene. Only until recently, the respective development of 3-D reconstruction from SAR and optical images allows a breakthrough in this research direction. As a first attempt, the "SARptical" system [1], [2] proposed a promising solution to tackle this challenging task.

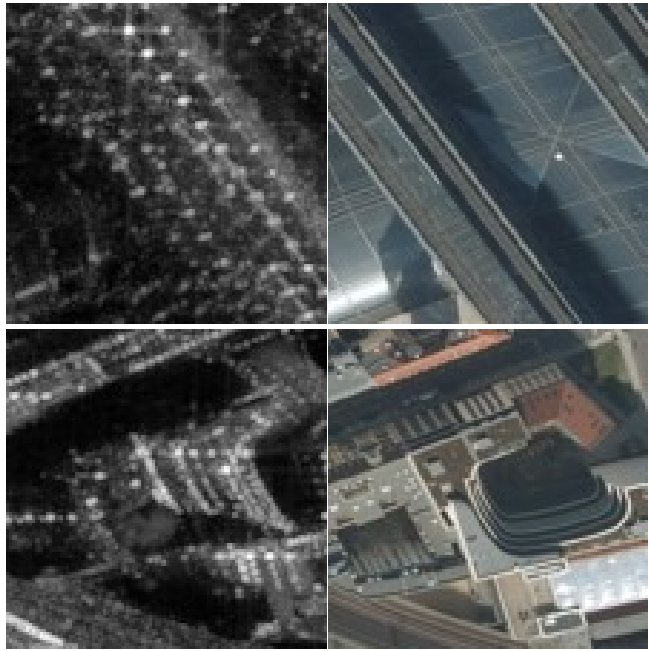

Figure 1. Sample SAR and optical patch pairs from the SARptical dataset. Left: SAR image patches, and right: the corresponding optical image patches. A corresponding SAR and optical patch pair refers to the 3-D positions of the center pixels of the two patches match within a margin of the accuracy of the 3-D reconstruction (typically a few meters).

The SARptical framework co-registers the 3-D models reconstructed from SAR and optical images respectively. As a result, the 2-D SAR and optical images are also matched. SARptical can trace individual SAR scatterers in the corresponding high resolution optical images where we can analyze the geometry, material, and other properties of the imaged objects. Vice versa, the similar study can also be done in the SAR image domain.

The goal of this paper is to introduce and share a dataset derived from the SARptical framework. The dataset consists of over 10,000 pairs of corresponding SAR and optical image patches extracted from TerraSAR-X high resolution spotlight images and aerial UltraCAM optical images. Figure 1 shows two examples of matched pairs of SAR and optical patches in the dataset. The left column is the SAR patches showing its amplitude in dB, and the right column is the corresponding optical patches. In the final version of this paper, we will make this dataset freely available for research. It can be downloaded from http://www.sipeo.bgu.tum.de/downloads/SARptical_data.zip .

## 2. THE SARPTICAL FRAMEWORK

The complex topography in dense urban areas causes inevitable layover and shadowing the SAR images. Hence, the core of SARptical is to match SAR and optical images in 3-D space. This requires additionally an accurate digital elevation model (DEM) of the area. In most occasions, such DEM is not available. Hence, SARptical also estimates the 3-D position of individual pixels from the SAR and optical images, respectively. The general workflow of SARptical is shown in Figure 2.

In order to estimate the 3-D position of the individual pixels in the images, the algorithm requires an interferometric stack of SAR images, as well as at least a pair of optical images. We estimate the 3-D point clouds from the SAR and the optical data using differential SAR tomographic inversion (D-TomoSAR) [4]–[8] and optical multi-view stereo matching, respectively. The matching of the two point clouds in 3-D guarantees the matching of the SAR and the optical images.

Finally, we can trace the coordinate of any SAR point in the optical image domain, or vice versa. This leads to a pair of corresponding SAR and optical patches whose center pixels' 3-D positions are "identical". Of course, the term "identical" is always subject to the accuracy of 3-D reconstruction of TomoSAR and stereo matching, respectively. This depends on the number of SAR and optical images, and their signal-to-noise ratio. For our dataset, the accuracy is in general about *one to two meters*.

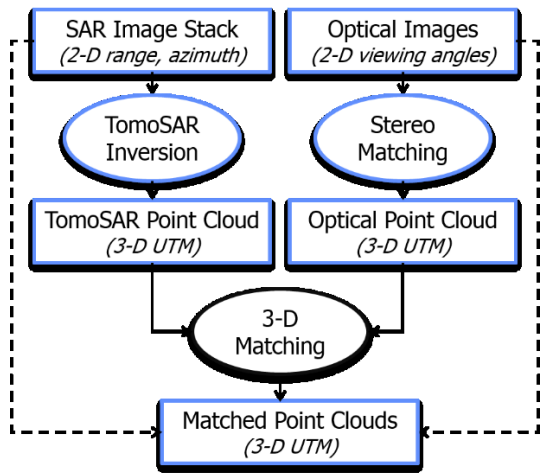

**Figure 2. Workflow of the SARptical framework. The coordinate system of each dataset in the flowchart is indicated by the italic text in the bracket. The dashed lines refer to that the SAR and optical images can be projected to each other through the matched 3-D point cloud.**

## 3. THE SARPTICAL DATASET

For this dataset, we made use of a stack of 109 TerraSAR-X high resolution spotlight images of Berlin acquired between 2009 and 2013 with about 1 meter resolution and 9 UltraCAM optical images of the same area with 20cm ground spacing [9]. D-TomoSAR and multi-view stereo matching were applied to the two datasets for 3-D reconstruction, respectively. After the 3D point cloud reconstruction, 32,446 pixels were selected from the SAR images and projected into the optical images, yielding 89,502 optical patches. Image patches of 112×112 pixels are centered at a given SAR pixel, and a similarly large patch around the projected position in the optical image is cropped to generate a pair of corresponding SAR-optical patches. The patches are about 100×100m ground coverage. Proper corrections, including rotation and adjustment of the pixel spacing, has been applied on the corresponding patches, so that they align with each other at a first approximation.

Figure 3 shows the co-registered SAR and optical point cloud overlaid with the two pairs of matched image patches shown in Figure 1. The left pair is on the roof of Berlin Central Station, and the right one is over a curved building next to the intersection of Reinhardtstraße and Luisenstraße.

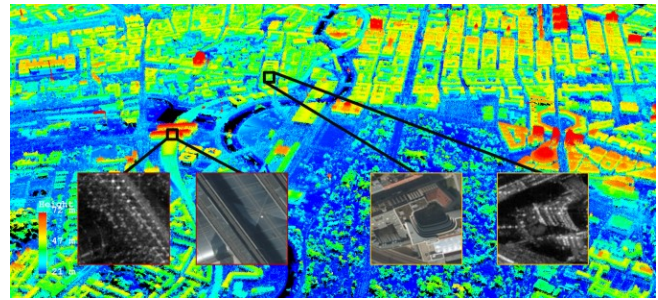

**Figure 3. A of schematic figure of the generation of the SARptical dataset. The background point cloud is the fusion of the TomoSAR point cloud and the optical point cloud. The color of the point cloud represents the height. The black rectangles mark the areas of the two examples shown in Figure 1.**

Because multiple optical images from different viewing angles were used in the 3-D reconstruciton, each SAR image patch may have maximum of nine corresponding optical image patches from different viewing angles, depending on the visibility of the SAR pixel from the respective optical point of view. Figure 4 shows two example of SAR patches that have multiple corresponding optical image patches from different viewing angles. As one can see that different façades are visible in different optical patches. This gives opportunity for new applications in urban monitoring.

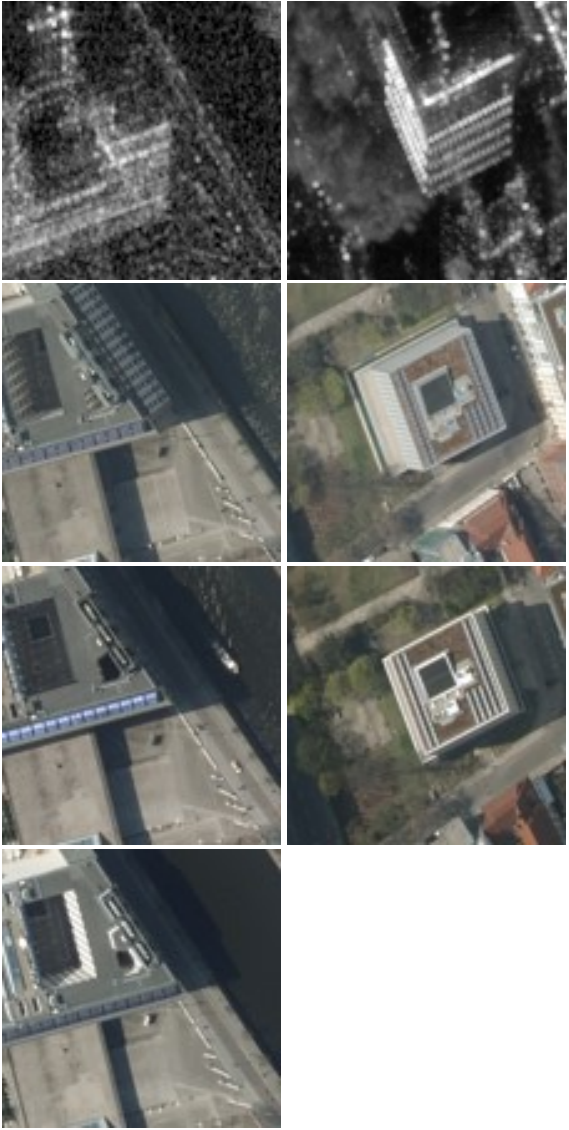

**Figure 4. Example of SAR image patches with multiple corresponding optical image patches from different view angles.**

## 4. CONCLUSION

This paper introduces and shares the SARptical dataset that is a database of over 10,000 pairs of matched very high resolution SAR and optical image patches. This dataset opens new opportunities for urban application, such as joint classification using SAR and optical images, as well as transfer learning between the two types of image. This dataset will be made available to the community. Please visit our website http://www.sipeo.bgu.tum.de/downloads for more details.

## 5. REFERENCES

[1] Y. Wang and X. X. Zhu, "Fusing Meter-Resolution 4-D InSAR Point Clouds and Optical Images for Semantic Urban Infrastructure Monitoring," *IEEE Trans. Geosci. Remote Sens.*, 2016.

[2] Y. Wang and X. X. Zhu, "InSAR Forensics: Tracing InSAR Scatterers in High Resolution Optical Image," presented at the Fringe 2015, 2015.

[3] L. Mou, M. Schmitt, Y. Wang, and X. X. Zhu, "A CNN for the identification of corresponding patches in SAR and optical imagery of urban scenes," in *2017 Joint Urban Remote Sensing Event (JURSE)*, 2017, pp. 1–4.

[4] X. Zhu and R. Bamler, "Very High Resolution Spaceborne SAR Tomography in Urban Environment," *IEEE Trans. Geosci. Remote Sens.*, vol. 48, no. 12, pp. 4296–4308, 2010.

[5] X. X. Zhu, Y. Wang, and R. Bamler, "Integration of Tomographic SAR Inversion and PSI for Operational Use," in *9th European Conference on Synthetic Aperture Radar, 2012. EUSAR*, 2012, pp. 151–154.

[6] Y. Wang, X. X. Zhu, R. Bamler, and S. Gernhardt, "Towards TerraSAR-X Street View: Creating City Point Cloud from Multi-Aspect Data Stacks," in *Urban Remote Sensing Event (JURSE), 2013 Joint*, 2013, pp. 198–201.

[7] X. Zhu, Y. Wang, S. Gernhardt, and R. Bamler, "Tomo-GENESIS: DLR's Tomographic SAR Processing System," in *Urban Remote Sensing Event (JURSE), 2013 Joint*, 2013, pp. 159–162.

[8] Y. Wang, X. Zhu, and R. Bamler, "An Efficient Tomographic Inversion Approach for Urban Mapping Using Meter Resolution SAR Image Stacks," *IEEE Geosci. Remote Sens. Lett.*, vol. 11, no. 7, pp. 1250–1254, 2014.

[9] Data provided by 'Land Berlin' and 'Business Location Service', supported by 'Europäischer Fonds für Regionale Entwicklung'